\documentclass[12pt,reqno]{amsart}
\usepackage{amsfonts}

\usepackage{amscd,amssymb,amsthm}

\newtheorem{theorem}{Theorem}[section]
\newtheorem{Lemma}[theorem]{Lemma}
\newtheorem{proposition}[theorem]{Proposition}

\newcommand{\Tr}{\mathop{\rm Tr}\nolimits}
\begin{document}
\author{M. Corgini and R. Tabilo}
\title[]{ Note on a Non Linear Perturbation of the Ideal Bose Gas}
\address{Universidad de La Serena\\
Cisternas 1200, La Serena. Chile}
\maketitle

\begin{abstract} In this work we show that the introduction of a $U(1)$ symmetry  breaking field  in the energy operator of the boson-free gas, is equivalent, in the thermodynamic limit, to the inclusion, in the Hamiltonian of the ideal gas, of a non-linear  function of the number operator associated with the zero mode. In other words, the limit pressures coincide. Moreover, both models undergo non  conventional Bose-Einstein condensation (BEC) for strictly negative values of the chemical potential $\mu.$  Finally, the proof of equivalence of limit pressures is extended to a class of full-diagonal models. 
\end{abstract}

\vspace{1cm}

\section{Introduction} 

Until 2013 it was believed that, from the point of view of a physical experiment, to confine a homogeneous system of Bose atoms, and to make it pass, subsequently, to the thermodynamic limit, would be an impossible task to perform. Thus \cite{MULL}, 

\vspace{0.5cm}

{\small{
\begin{quotation} In the magnetic traps, not only is the number of particles quite small, compared to the usual case, but the ``bounda\-ry,'' formed by a quadratic potential well, extends literally throughout the whole system. In order to take the thermodynamic limit in such a system it is necessary to weaken the potential so that, as the number of particles increases, the average density remains constant. This is well-defined mathematically, but is of course physically unrealizable. On the other hand, ta\-king the box size to infinity in the homogeneous case is also unrealized experimentally.
\end{quotation}}} 

\vspace{0.5cm}

Moreover, trapped gases were, generally speaking,  spatially inhomogeneous. In this framework, to overcome the difficulties in defining pressure and volume for a gas confined in an inhomogeneous trap, it has been necessary to define macroscopic parameters that behave like them.

However, in 2013,  BEC in a quasi uniform three dimensional potential of an optical trap box (cilindrical optical box) BEC was observed \cite{GAUNT,GAUNT2}. 

The authors of ref.\cite{GAUNT} point out that:

\vspace{0.5cm}

{\small{
\begin{quotation}We have observed the Bose-Einstein condensation of an atomic gas in the (quasi)uniform three-dimensional potential of an optical box trap. Condensation is seen in the bimodal momentum distribution and the anisotropic time-of-flight expansion of the condensate. The critical temperature agrees with the theoretical prediction for a uniform Bose gas. The momentum distribution of a noncondensed quantum-degenerate gas is also clearly distinct from the conventional case of a harmonically trapped sample and close to the expected distribution in a uniform system. We confirm the coherence of our condensate in a matter-wave interference experiment. Our experiments open many new possibilities for fundamental studies of many-body physics.
\end{quotation}}}

\vspace{0.5cm}

Later, in the same article, they indicate that,

\vspace{0.5cm}

{\small{
\begin{quotation}The thermodynamics of our gas are
therefore very close to the textbook case of a uniform
system and very different from the case of a harmonically
trapped sample.
\end{quotation}}}

\vspace{0.5cm}

In this sense, this seems to be a suitable experimental scenario to test the consistency of  the Bogoliubov's theory -based on the concept of \emph{quasiaverages} ~\cite{BOG}-, about the spontaneous rupture of the $U(1)$ symmetry and simultaneous emergence of Bose Einstein condensation in the case of an ideal Bose gas. In other words, experiments could be carried out, in this framework, to study the thermodynamic behavior of an Ideal Bose gas system when it is disturbed by a nonvanishing external field,  that breaks the symmetry $U (1)$. The question is: what is, in this case, the nature of such a condensation?

\vspace{0.5cm}

\section{Basic Notions}

\subsection{Grand canonical and canonical ensembles}

Let $ \hat{H}_l $  be a selfadjoint operator  on the Hilbert space ${\mathcal{F}}_{\mathrm B}$ (Fock space),  representing the energy operator of  a Bose particle system. Let $\beta = T^{-1}$ and $\mu \in \mathbb{R}$ be  the inverse temperature and the so-called  chemical potential, respectively.  

$\hat{H}_l (\mu )$ is defined as $ \hat{H}_l (\mu ) :=\hat{H}_l  -\mu \hat{N},$ where $\hat{N}$ is the total number operator given by $\hat{N} = \displaystyle \sum_{{\mathbf{p}}} \hat{n}_{\mathbf{p}},$ being $ \hat{n}_{\mathbf{p}} = \hat{a}^{\dag}_{\mathbf{p}}\hat{a}_{\mathbf{p}}$ the number operator associated to the ${\mathbf{p}}-$ mode.

The operators $\hat{a}^{\dag}_{\mathbf{p}},\hat{a}_{\mathbf{p}}$ defined on the Fock space,  well-known as creations and  annihilation operators, respectively, satisfies the commutation rules:
$$ [\hat{a}_{\mathbf{p}}, \hat{a}^{\dag}_{\mathbf{q}} ] = \delta_{\mathbf{p}\mathbf{q}}I, $$ being $\delta_{\mathbf{p}\mathbf{q}}$ the kronecker delta and $I$ the identity operator.

$\hat{H}_l $ can be decomposed as the followig sum: $\hat{H}_l  =  \hat{H}^0_l + \hat{H}^I_l, $ where  $\hat{H}^0_l= \displaystyle \sum_{\mathbf{p}} \lambda_l ({\mathbf{p}}) \hat{a}^{\dag}_{\mathbf{p}} \hat{a}_{\mathbf{p}} $  and 
$ \hat{H}^I_l= \displaystyle \sum_{{\mathbf{p}},{\mathbf{q}},{\mathbf{r}},{\mathbf{s}}}  U_{{\mathbf{p}},{\mathbf{q}},{\mathbf{r}},{\mathbf{s}}}   \hat{a}^{\dag}_{\mathbf{p}}\hat{a}^{\dag}_{\mathbf{q}} \hat{a}_{\mathbf{r}}\hat{a}_{\mathbf{s}}$ are the seccond quantizations of the laplacian operator and of the interaction $U,$ respectively, both defined on the region of confinement of particles $\Lambda \subset \mathbb{R}^d,$ with $ d\in \mathbb{N}.$ 

We shall assume in this work, periodic boundary conditions. In this case all the subscripts 
$\mathbf{p}$ belong to  the set $\Lambda^*$ (dual of $\Lambda$) defined as $\Lambda^*_l = \{ \mathbf{p}= (p_1,\dots,p_{d})\in {\mathbb R}^{d}:
p_{\alpha}={2\pi n_{\alpha}}/{l}, n_{\alpha} \in {\mathbb Z}, \alpha
=1,2,...,d\},$ and $\lambda_{l}(\mathbf{p})= {\mathbf{p}^2}/{2}.$  

\smallskip 

With these definitions, at finite volume,  it is posible to introduce the grand canonical partition function $ \Xi_l (\beta,\mu) $ and the pressure $p_{l}(\beta,\mu ):$ 
$$ \Xi_l (\beta,\mu) := \mathop{\rm Tr}\nolimits_{%
\mathcal{F}_B} \exp \left( -\beta \hat{H}_l (\mu)\right),\;p_{l}(\beta,\mu ) :=\frac{1}{\beta V_l} \ln \Xi_l (\beta,\mu);
  $$  the canonical partition function $Z_{N,l} (\beta,\varrho)$ and the free energy
$f_{l}(\beta ,\varrho_l ), $ where $ \varrho_l = \frac{N}{V_l}:$ 

$$ Z_{N,l} (\beta,\varrho) := {\rm Tr}_{{\mathcal{H}}_{B}^{(N)}}e^{-\beta {\hat{H_l}}^{(N)}},\;\;f_{l}(\beta ,\rho_l ) := -\frac{1}{\beta V_l} \ln  Z_{N,l} (\beta,\varrho), $$ and, finally, the Gibbs states in the grand canonical ensemble and in the canonical ensemble:

$$\langle \cdot \rangle_{\hat{H}_l(\mu)} = \Xi^{-1} (\beta,\mu)
\mathop{\rm Tr}\nolimits_{\mathcal{F}_B} \cdot \exp \left( -\beta \hat{H}
_l (\mu)\right), $$ 

$$\langle \cdot \rangle_{\hat{H}^{(N)}_l}= Z_{N,l} (\beta,\varrho)
\mathop{\rm Tr}\nolimits_{ \mathcal{H}^{(N)}_B} \cdot \exp \left( -\beta \hat{H}^{(N)}
_l \right), $$ respectively.

The limit free energy $ f (\beta, \varrho)$ and the limit pressure $ p (\beta, \mu)$ are defined as:

$$f (\beta, \varrho) :=\displaystyle\lim_{N_l, V_l\to\infty} f_l(\beta,\varrho_l),\;\mbox{assuming that} \displaystyle\lim_{N_l,V_l\to\infty}\varrho_l = \varrho = \mbox{constant},$$ and

$$p (\beta, \mu):= \displaystyle\lim_{V_l\to\infty} p_l(\beta,\mu),\;\;\mbox{with} $$ $$ \displaystyle\lim_{V_l\to\infty}\left\langle \frac{\hat{N}}{V_l}\right\rangle_{\hat{H}%
_l (\mu)} = \displaystyle\lim_{V_l\to\infty}\rho_{l} = \rho (\mu) =  \mbox{constant}.$$

On the other hand, stable systems are defined as those for which there exists $\mu_* \in \mathbb{R}$ such that only for $\mu\in (-\infty, \mu_*],$  $p (\beta, \mu) < \infty,$ while superstable sytems satisfies $p (\beta, \mu) < \infty.$ for all values of $\mu.$ Finally, if the following inequality (in the sense of operators)

$$
\hat{H}^I_l \geq -\frac{C_2}{V_l}\hat{N} + \frac{C_1}{V_l}\hat{N}^2
$$ holds, the system is superstable. 

\bigskip

\subsection{Types of BEC}
{
\begin{itemize}
\item Condensation of type I corresponds to a macroscopic occupation of a finite nuber of states. Thus, a macroscopic occupation of the the ground state, or traditional Bose-Einstein condensation,  is given by the fulfilment of the condition $$ \displaystyle\lim_{V_l\to \infty}\left< \frac{\hat{n}_{\mathbf{0}}}{V_l} \right>_{ {\hat{H}_l(\mu)}} = \rho_{\mathbf{0}} > 0.$$ For the latter, in the condensed-uncondensed phase transition the appropiate order parameter is $\rho_{\mathbf{0}} = \rho -\rho_c,$ being $\rho_c$ a critical density. 

\item Condensation of type II  holds when there exists an infinite number of states macoscopically occupied.

\item Condensation of the type III holds when there are not macroscopically occu\-pied states but the following condition holds: 

$$
\displaystyle\lim_{\delta\to 0^+}\lim_ {V_l\to\infty} \frac{1}{V_l}
\displaystyle\sum_{\mathbf{p}\in \Lambda^*, \lambda_l (\mathbf{p}) < \delta }\langle
\hat{n}_{\mathbf{p}}\rangle_{\hat{H}_l(\mu)}> 0.
$$
\end{itemize}
}

The third type of Bose condensation, denominated \emph{generalized BEC} (GBEC), was introduced by M. Girardeau in 1960~\cite{GIR}. 

GBEC is more robust that the other kinds of condensation in the sense that it is independent on the shape of the confining region. Indeed, in the case of the free Bose gas, it always occurs for particle density values larger than a critical one. 

These kinds of critical phenomena are in agreement with the standard phase transitions theory that identifies critical points with the emergence of singularities in the thermodynamic functions in the thermodynamic limit. 

 However, there is a fourth type of condensation independent on temperature and, for this reason, called non conventional. It is in the study of this phenomenon that we are interested in this work.

\bigskip

\section{BEC and spontaneous symmetry breaking (SSB)}

The standard strategy devoted to associate symmetry breaking with certain phase transition consists in introducing a small term on the original energy operator, preserving its self-adjointness  but eliminating the symmetry corresponding to some conservation law. 

Thus, in the case of Bose systems, it is posible to break the global  $U(1)$ symmetry by adding the extra term $ -\sqrt{V}\nu \left( \hat{a}_{\mathbf{0}}e^{-i\varphi} +  \hat{a}^{\dag}_{\mathbf{0}}e^{i\varphi}\right) $  to the original energy operator $ \hat{H}_{l} (\mu)=\hat{H}_{l} - \mu\hat{N},$ which satisfies $ [\hat{H}_{l} (\mu), \hat{N}],$  being $\hat{N}$ the total number operator, obtaining  the new Hamiltonian $ \hat{H}_{l,\nu,\varphi} (\mu)=  \hat{H}_l (\mu) -\sqrt{V}\nu \left( \hat{a}_{\mathbf{0}}e^{-i\varphi} +  \hat{a}^{\dag}_{\mathbf{0}}e^{i\varphi}\right)$ for which $ [\hat{H}_{l,\nu,\varphi} (\mu),\hat{N} ] \neq 0,$ being $\nu\in \mathbb{R}^+,\;\varphi \in [0,2\pi).$ In this case, the typical selection rules (degeneracy of the thermal averages), 

$$ \left|\left< \hat{a}^{\dag}_{\mathbf{0}}  \right>_{\hat{H}_{l}(\mu)}\right| = \left| \left<\hat{a}_{\mathbf{0}}  \right>_{\hat{H}_{l}(\mu)}\right| = 0, $$ being $\left< - \right>_{\hat{H}_{l}(\mu)} $ the thermal average associated to  $\hat{H}_{l} (\mu),$ at finite volume $V= l^d,\; d\in \mathbb{N},\; d\geq 3,$  do not hold anymore, i.e.:

$$ \left|\left< \hat{a}^{\dag}_{\mathbf{0}}  \right>_{\hat{H}_{l,\nu,\varphi}(\mu)}\right| = \left| \left<\hat{a}_{\mathbf{0}}  \right>_{\hat{H}_{l,\nu, \varphi}(\mu)}\right| = \sqrt{V}\eta_{l,\nu,\varphi} \neq 0, $$ where $\left< - \right>_{\hat{H}_{l,\nu, \varphi}(\mu)} $ is the thermal average corresponding to the perturbed operator  $\hat{H}_{l,\nu,\varphi} (\mu).$

For a Bose system undergoing BEC, in the thermodynamic limit, we have

$$\displaystyle\lim_{\nu \to 0} \displaystyle\lim_{V\to \infty}\eta^2_l= \left\{
\begin{array}{cl}
\rho_{\mathbf{0}} \neq 0 &\mbox{if } \rho > \rho_c\\
0 &\mbox{if } \rho \leq \rho_c
\end{array}\right.$$ being $\rho_c $ a critical density of particles.

From a mathematical point of view, for $\rho \leq \rho_c,$ in the uncondensed phase,  it is possible to make a  limit exchange, obtaining:  $$\displaystyle\lim_{\nu \to 0} \displaystyle\lim_{V_l \to \infty}\eta_{l,\nu,\varphi} =   \displaystyle\lim_{V\to \infty}\displaystyle\lim_{\nu \to 0} \eta_{l,\nu,\varphi} = 0.$$  However, for $\rho > \rho_c,$  $$\displaystyle\lim_{\nu \to 0} \displaystyle\lim_{V_l \to \infty} \eta_{l,\nu,\varphi}\neq   \displaystyle\lim_{V\to \infty}\displaystyle\lim_{\nu \to 0} \eta_{l,\nu,\varphi}$$

In this context the limit thermal averages defined as $$ \prec \cdot \succ := \displaystyle\lim_{h\to 0}\lim_{V\to\infty}\left< - \right>_{\hat{H}_{l,\nu,\varphi}(\mu)}$$
 have been  denominated {\it Bogoliubov's quasiaverages} or {\it anomalous ave\-rages}. In fact, this notion was introduced for the first time by  N.N.Bogo\-liubov~\cite{BOG}.

Thus, the degeneracy of regular averages, produced by the presence of additive conservation laws (or equivalently, by the invariance of the Hamiltonian with respect to certain groups of transformations)  is reflected by the dependence of quasi averages on the extra infinitesimal term. In this sense Bogoliubov claimed that the latter are more ``physical'' than the regular averages \cite{BOG}. However this procedure, in some cases, has been applied \emph{without having necessarily a clear physical meaning}.

\bigskip

{\small{
\begin{quotation}
We are assuming that other types of degeneracy do not exist and, thus, the introduction of the term [...] is suffi\-cient for the removal of the degeneracy.
(N. N. Bogoliu\-bov \cite{BOG})
\end{quotation}}}

\bigskip

Let $\hat{\rho}_{\mathbf{0},l} =  V^{-1}{\hat{a}^{\dag}_{\mathbf{0}}\hat{a}_{\mathbf{0}}},$ $\hat{\eta}_l = {V}^{-\frac{1}{2}}{\hat{a}_{\mathbf{0}}}.$ In the case of the free Bose gas,  for 

\begin{equation}\label{eq1} \hat{H}^{0}_{l,\nu,\varphi} = \displaystyle\sum _{\mathbf{p}} \lambda_l (\mathbf{p}) \hat{a}^{\dag}_{\mathbf{p}}\hat{a}_{\mathbf{p}}-\sqrt{V}\nu \left( \hat{a}_{\mathbf{0}}e^{-i\varphi} +  \hat{a}^{\dag}_{\mathbf{0}}e^{i\varphi}\right),\end{equation} 

\bigskip

\noindent
the following limits

\[\displaystyle\lim_{\nu\to 0^+} \displaystyle\lim_{V\to \infty} \left< \hat{\rho}_{\mathbf{0},l}  \right>_{\hat{H}_{l, \nu, \varphi}(\mu)}  = \rho_{\mathbf{0}},\;\;\displaystyle\lim_{\nu\to 0^+}\displaystyle\lim_{V\to \infty}\left< \hat{\eta}_l  \right>_{\hat{H}_{l, \nu, \varphi}(\mu)} = \sqrt{\rho_{\mathbf{0}}}e^{i\varphi } \] hold \cite{BOG}. In other words: $$ \displaystyle\lim_{\nu\to 0^+} \displaystyle\lim_{V\to \infty} \left< \hat{\rho}_{\mathbf{0},l}  \right>_{\hat{H}_{l, \nu, \varphi}(\mu)}   = \displaystyle\lim_{h\to 0^+}\displaystyle\lim_{V\to \infty}|\left< \hat{\eta}_l  \right>_{\hat{H}_{l, \nu, \varphi}(\mu)} |^2= \rho_{\mathbf{0}}.$$

\bigskip

In what follows, in order to simplify the notation, we will omit the angle $\varphi$ in the subscripts of any mathematical expression (thermal ave\-rages, Hamiltonians, etc.).

\bigskip

Following step by step the strategy developed by Bogoliubov, let

$$
\hat{a}_{\mathbf{0}} = -\frac{\nu}{\mu} e^{i\varphi}\sqrt{V} + \hat{b}_{\mathbf{0}},\;\;
\hat{a}^{\dag}_{\mathbf{0}} = -\frac{\nu}{\mu} e^{-i\varphi}\sqrt{V} + \hat{b}^{\dag}_{\mathbf{0}}.
$$

Substituting this operators in the original energy operator, we obtain:
$$  \hat{H}^0_{l,\nu} (\mu) = -\mu \hat{b}^{\dag}_{\mathbf{0}}\hat{b}_{\mathbf{0}} +  \displaystyle\sum_{\mathbf{p}\in {\Lambda_l}^*\backslash\{0\}} \left(\frac{\mathbf{p}^2}{2}-\mu\right) \hat{a}^{\dag}_{\mathbf{p}}\hat{a}_{\mathbf{p}}  + \frac{\nu^2 V}{\mu}.$$ 

In that follows, it will be assumed that,
$$\mu = \mu_* = -\frac{\nu}{\sqrt{\rho_{0}}},$$ where $\rho_0 $ is a strictly positive real constant. 

Clearly, $$\left<   \hat{b}^{\dag}_{\mathbf{0}}   \right>_{\hat{H}^0_{l,\nu}( \mu_* )} = \left<   \hat{b}_{\mathbf{0}}   \right>_{\hat{H}^0_{l,\nu}( \mu_* )}= 0.$$  This implies that:

$$  \left<   \hat{a}^{\dag}_{\mathbf{0}}   \right>_{\hat{H}^0_{l,\nu}( \mu_* )} =
 e^{-i\varphi}\sqrt{V\rho_{\mathbf{0}}},\;\;  \left<   \hat{a}_{\mathbf{0}}   \right>_{\hat{H}^0_{l,\nu}( \mu_* )} =
 e^{i\varphi}\sqrt{V\rho_{\mathbf{0}}} $$

\bigskip

Besides,

$$ \left<   \hat{b}^{\dag}_{\mathbf{0}}\hat{b}_{\mathbf{0}}   \right>_{\hat{H}^0_{l,\nu}( \mu_* )}  = \left(\exp \beta \left(- \mu_*  \right)-1\right)^{-1},\;\;\left<   \hat{n}_{\mathbf{p}}   \right>_{\hat{H}^0_{l,\nu}( \mu_* )}=   \left(\exp \beta \left(\frac{\mathbf{p}^2}{2}- \mu_* \right)-1\right)^{-1}. $$ 

Then, we define

$$  \rho_{c,l} (\beta,  \mu_* )=  \frac{1}{V}  \displaystyle\sum_{\mathbf{p}\in {\Lambda_l}^*\backslash\{0\}} \left(\exp \beta \left( \frac{\mathbf{p}^2}{2} - \mu_*  \right)-1\right)^{-1}.$$ Passing to the thermodynamic limit, we get:

$$
 \rho_c (\beta, \mu_* )=   \frac{1}{(2\pi)^d}  \int \left(\exp \beta \left(\frac{\mathbf{p}^2}{2} -  \mu_* \right)-1\right)^{-1}d^3\mathbf{p}.
$$

On the other hand,

$$
\left<   \frac{\hat{b}^{\dag}_{\mathbf{0}}\hat{b}_{\mathbf{0}}}{V}\right>_{\hat{H}^0_{l,\nu}( \mu_* )}  = \left< \left(\frac{\hat{a}^{\dag}_{\mathbf{0}}}{\sqrt{V}}  -  \sqrt{\rho_{\mathbf{0}}} e^{-i\varphi}  \right)\left(\frac{\hat{a}_{\mathbf{0}}}{\sqrt{V}}  -  \sqrt{\rho_{\mathbf{0}}} e^{i\varphi}  \right)\right>_{\hat{H}^0_{l,\nu}( \mu_* )} 
$$

$$
= \displaystyle\lim_{V\to\infty} \frac{1}{V} \left(\exp \beta \left(  -\mu_*  \right)-1\right)^{-1} = 0.
$$

This leads to,

$$
\displaystyle\lim_{V\to\infty} \frac{1}{V} (\left<   \hat{a}^{\dag}_{\mathbf{0}}\hat{a}_{\mathbf{0}}   \right>_{\hat{H}^0_{l,\nu}( \mu_* )} - \sqrt{\rho_{\mathbf{0}}}e^{i\varphi}  \left<   \frac{\hat{a}^{\dag}_{\mathbf{0}}}{\sqrt{V}} \right>_{\hat{H}^0_{l,\nu}( \mu_* )}  $$ $$  - \sqrt{\rho_{\mathbf{0}}}e^{-i\varphi}  \left<   \frac{\hat{a}_{\mathbf{0}}}{\sqrt{V}} \right>_{\hat{H}^0_{l,\nu}( \mu_* )} + \rho_{\mathbf{0}})= 0.
$$

$$
\displaystyle\lim_{V\to\infty} \frac{1}{V} \left<   \hat{a}^{\dag}_{\mathbf{0}}\hat{a}_{\mathbf{0}}   \right>_{\hat{H}^0_{l,\nu}( \mu_* )} = \displaystyle\displaystyle\lim_{V\to\infty} 
\left| \left<   \frac{\hat{a}^{\dag}_{\mathbf{0}}}{\sqrt{V}}\right>_{\hat{H}^0_{l,\nu}( \mu_* )} \right|^2 = \rho_{\mathbf{0}}.
$$

\bigskip

The role of the coupled external source, which is very unique, should not be exaggerated from a physical point of view. It is rather the superposition or transition from the ground state to a coherent state that best reflects, in that sense, the spontaneous break of symmetry. Such superposition disappears when the thermodynamic limit is reached. There, both states (with the same energy) the fundamental and the coherent - traslation of the first one- become orthogonal between them. Mathenati\-cally speaking, unlike the finite systems, in the thermodynamic limit we have infinite inequivalent representations of the Bose commutation rules or, in other words, infinite representations of the broken symmetry. In this sense, the Bogoliubov's approach consists in explicitely fixing one of them.

\bigskip

\section{Non linear perturbation of the Ideal Bose Gas }

Our purpose, in this work,  is to replace in eq.(1) the external source   $ -\sqrt{V}\nu \left( \hat{a}_{\mathbf{0}}e^{-i\varphi} +  \hat{a}^{\dag}_{\mathbf{0}}e^{i\varphi}\right) $  by $-2\nu\sqrt{V}\sqrt{ \hat{n}_{\mathbf{0}} + 1  }.$ In this case, the latter exxpression is expandable in powers series of the operator $ \hat{n}_{\mathbf{0}}$ (spectral theorem). This substitution is motivated by the fact that:

$$
\hat{a}_{\mathbf{0}} \varphi_n = \sqrt{n_{\mathbf{0}}} \varphi_{n-1},\;\hat{a}^{\dag}_{\mathbf{0}} \varphi_n = \sqrt{n_{\mathbf{0}}+ 1} \varphi_{n + 1},
$$

\bigskip

\noindent
where $\{ \varphi_{n}\} $ is a set of orthonormal eigenfunctions of $\hat{n}_{\bf{0}}.$ 

\bigskip
Thus, in this section  we consider a  model of a Bose gas whose energy operator correponds to the sum of the Hamiltonian of the free Bose gas with a nonlinear perturbation represented by the square root of the number operator  associated to the zero mode.

\begin{equation}\label{eq2} \hat{H}^{\rm{app}}_{l,\nu} (\mu)= \displaystyle\sum_{\mathbf{p}\in {\Lambda_l}^*} \lambda_l(\mathbf{p}) 
\hat{a}^{\dag}_{\mathbf{p}}\hat{a}_{\mathbf{p}} -  2\nu\sqrt{V}\sqrt{ \hat{n}_{\mathbf{0}} + 1  }  -\mu\hat{N},\;\mbox{where}\;\nu >  0, \end{equation} 

\bigskip

\noindent
 The Hamiltonian given by eq.(\ref{eq2}) represents a stable model defined in the domain $\mathcal{D} = \{(\beta,\mu): \beta > 0,\; \mu < 0\}. $  

On the other hand, let $\hat{H}^{0}_{l,\nu}(\mu)$ is defined as: 

\bigskip

\begin{equation}\label{eq3}\hat{H}^{0}_{l,\nu}(\mu) = \hat{H}^{0}_{l}(\mu) -\nu\sqrt{V} (e^{i\varphi}\hat{a}^\dag_{\mathbf{0}} +   e^{-i\varphi}\hat{a}_{\mathbf{0}}),\;\nu >0.\end{equation}

\bigskip

Note that $ [ \hat{H}^{\rm{app}}_{l,\nu} (\mu),\hat{N}] = 0, $ i.e. the energy operator given by eq.(\ref{eq2})  preserves the $U(1)$ symmetry. However,
$ [\hat{H}^{0}_{l,\nu}(\mu),\hat{N}] \neq 0,$ i.e., the continuous gauge symmetry associated with the $U(1)$ group is broken by the external field 
$ -\nu\sqrt{V} (e^{i\varphi}\hat{a}^\dag_{\mathbf{0}} +   e^{-i\varphi}\hat{a}_{\mathbf{0}}).$

In the next section a strong connection between the critical behaviour of both models, in the thermodynamic limit,  will be stablished. 

The main purpose of this work is to determine explicit expressions for the limit pressures of the model given by eq.(\ref{eq2}) in the framework of the so called  Laplace principle (see Apendix)  and the Large Deviations Method based in two theorems proved by  S. R. S. Varadhan ~\cite{VAR}. Moreover, it  shall be proven the existence of a phase characrterized by the emergence of non conventional Bose-Einstein condensation, i.e., the existence of an independent on tempe\-rature condensate.

\bigskip

\subsection{Limit pressure and nonconventional condensation}

\bigskip

\begin{theorem}\label{thm1} For $ (\beta,\mu ) \in \mathcal{D},\;\nu > 0,$  in the thermodynamic limit, 
\begin{equation}\label{eq4} p^{\rm{app}} (\beta,\mu,\nu)  = -\frac{\nu^2}{\mu}+  p^{\rm{id}^{\prime}} (\beta,\mu),\end{equation} where  $ p^{\rm{app}} (\beta,\mu,\nu), p^{\rm{id}^  \prime} (\beta,\mu)$  are the the limit pressures of the system whose Hamiltonian is given by the energy operator of eq.(\ref{eq2})  and  the energy operator given by  eq.(\ref{eq1}), but excluding  the mode $\mathbf{0},$ respectively.
\end{theorem}

\bigskip

\begin{proof}

Let,

\bigskip

$$\hat{H}^{\rm{app}}_{l,\nu} = \displaystyle\sum_{\mathbf{p}\in \Lambda^*} \lambda ({\mathbf{p}})\hat{n}_{\mathbf{p}} - 2\sqrt{V}\nu\sqrt{\hat{n}_{\mathbf{0}}+1}.$$ 

\bigskip

Note that the function $h(x) = ax + b\sqrt{x+ c},\;a,\;b\in \mathbb{R}, \;x \in [0,\infty ),\; c> 0$ is either an infinitely increasing or an infinitely decreasing mapping on $ [0, \infty)$ except the case  $ a <0,\; b>0.$ 

\bigskip

\noindent
Let $\{g_l\}$ be a sequence of functions defined on $[0,\infty)$ given as,

$$g_l(x) = (\mu-\lambda (\mathbf{0}))x
 + 2\nu\sqrt{ x+\frac{1}{V}},\; x \in [0,\infty),$$ whose first and second derivatives are, 

\bigskip

$$g^{\prime}_l(x) = (\mu-\lambda (\mathbf{0})) + \nu  \left( x + \frac{1}{V} \right)^{-1/2}, $$

$$g^{\prime\prime}_l (x) = -\frac{\nu}{2} \left( x + \frac{1}{V} \right)^{-3/2} < 0,$$ respectively.

From these facts, it follows that $g_l(x) $ is a concave function attaining its global  maximum at 

$$ x^*_l = \left( \frac{\nu}{\lambda ((\mathbf{0}) -\mu} \right)^2 -\frac{1}{V},$$ for a large enough value of $V$ such that $ x^*_l \geq 0$ and

$$\displaystyle\lim_{V\to\infty}\sup_{x\in [0,\infty)} g_l(x) = \displaystyle\lim_{V\to\infty}\sup_{x\in [0,\infty)} g_l(x^*_l)  = -\frac{\nu^2}{\mu},$$ being
$\mu < 0,$ $\lambda (\mathbf{0}) = 0.$ 

Use will be made  of the so-called  large deviations method, based on the Laplace principle, for obtaining a closed analytical expression for $ p^{\rm{app}} (\beta,\mu, \nu).$ Since $\hat{H}^{\rm{app}}_{l,\nu}(\mu) $ is a diagonal operator with respect to the number opera\-tors, the finite pressure can be written as,

$$  p^{\rm{app}}_l (\beta,\mu, \nu) =\frac{1}{\beta V}\ln \Tr _{\mathcal{F}_B} \exp \{ -\beta\hat{H}^{\rm{app}}
_{l,\nu}(\mu) \}$$

$$
 p^{\rm{app}}_l (\beta,\mu, \nu)_l = \frac{1}{\beta V}\ln\displaystyle\sum^{\infty}_{n_{\mathbf{0}} = 0} \exp \beta\{ (\mu -\lambda (\mathbf{0})) n_{\mathbf{0}} + 2\sqrt{V}\nu\sqrt{ n_{\mathbf{0}}+1}\}
$$

$$
+ \frac{1}{\beta V}\ln \displaystyle\sum^{\infty}_{\mathbf{p}\in\lambda^*\backslash \{\mathbf{0}\}, n_{\mathbf{p}}}\exp \beta\{ (\mu-\lambda (\mathbf{p}))n_{\mathbf{p}}\}.
$$

Noting, that

$$
p^{\rm{app, \mathbf{0}}}_l (\beta,\mu, \nu)_l=\frac{1}{\beta V}\ln\displaystyle\sum^{\infty}_{n_{\mathbf{0}} = 0} \exp \beta  \{ (\mu -\lambda (\mathbf{0})) n_{\mathbf{0}} + 2\sqrt{V}\nu\sqrt{ n_{\mathbf{0}}+1}\}
$$

$$
= \frac{1}{\beta V}\ln\displaystyle\sum^{\infty}_{n_{\mathbf{0}} = 0} \exp \beta V \left\{ (\mu -\lambda (\mathbf{0})) \frac{n_{\mathbf{0}}}{V} + 2\nu\sqrt{ \frac{n_{\mathbf{0}}}{V}+ \frac{1}{V}}\right\}
$$

$$
= \frac{1}{\beta V}\ln\displaystyle\sum^{\infty}_{n_{\mathbf{0}} = 0} \exp \left\{\beta Vg_l \left(\frac{n_\mathbf{0}}{V}\right)\right\}. $$ It is not hard to see that $\{   p^{\rm{app},\mathbf{0}}_l (\beta,\mu, \nu)  \}$ is a sequence of Darboux sums, then, in the thermodynamic limit the Laplace principle leads to the following expression,

$$ p^{\rm{app}} (\beta,\mu, \nu) =  -\frac{\nu^2}{\mu} +  p^{\rm{id}^\prime} (\beta,\mu). $$

\end{proof}

\bigskip

\begin{theorem}\label{thm2} For $ (\beta,\mu ) \in \mathcal{D},\;\nu > 0,$  in the thermodybanic limit, the Bose Gas with Hamiltonian given by eq.(2) undergoes non conventional condensation if and only if, the ideal gas whose energy operator is given by eq.(1) also displays  independent on temperature condensation.  Moreover, 

\bigskip

\begin{equation}\label{eq5}
p^{\rm{id}} (\beta,\mu,\nu) = p^{\rm{app}} (\beta,\mu,\nu), 
\end{equation} and the amount of condensate satisfies:

\bigskip

\begin{equation}\label{eq6} \rho^{\rm{app}}_{\mathbf{0}} (\mu, \nu) = \rho^{\rm{id}}_{\mathbf{0}} (\mu,\nu) =
 \frac{\nu^2}{\mu^2} \end{equation}
\end{theorem}

\bigskip

\begin{proof}

Note that:

\bigskip

\begin{equation}\label{eq7} p^{\rm{id}}_l (\beta,\mu,\nu) = \frac{1}{\beta V}\ln \left( 1-e^{\beta \mu}
 \right) - \frac{\nu^2}{\mu} +  p^{\rm{id}^\prime}_l (\beta,\mu).\end{equation} 

\bigskip

From this it follows that: 

\bigskip

$$ p^{\rm{id}}_l (\beta,\mu,\nu) - p^{\rm{app}} (\beta,\mu,\nu)  = \frac{1}{\beta V}\ln \left( 1-e^{\beta \mu} 
 \right) + p^{\rm{id}^\prime}_l (\beta,\mu) - p^{\rm{id}^\prime} (\beta,\mu). $$ 

\bigskip

\noindent
Thus, in the thermodynnamic limit, for  fixed values of $\beta $ y $\mu < 0,$

\bigskip

 $$p^{\rm{id}} (\beta,\mu,\nu) = p^{\rm{app}} (\beta,\mu,\nu).$$ 

\bigskip

On the other hand, using the Griffiths Lemma ~\cite{GRI}

\bigskip

$$\partial_{\mu} p^{\rm{id}}_l (\beta,\mu,\nu) - \partial_{\mu} p^{\rm{app}} (\beta,\mu,\nu) =\rho^{\rm{id}}_l (\beta,\mu,\nu) - \rho^{\rm{app}} (\beta,\mu,\nu)$$ 
$$= \frac{1}{V}\left(\frac{1}{e^{-\beta\mu}-1}\right).$$  In this case, 

\bigskip

\begin{equation}\label{eq8} \rho^{\rm{app}} (\beta,\mu, \nu)= \frac{\nu^2}{\mu^2} + \rho_{c}(\beta,\mu),\end{equation}

\begin{equation}\label{eq9} \rho^{\rm{id}}_l (\beta,\mu, \nu)= \frac{\nu^2}{\mu^2} +  \frac{1}{V}\left(\frac{1}{e^{-\beta\mu}-1}\right) + \rho_{c,l}(\beta,\mu).\end{equation}

\bigskip

From eqs.(\ref{eq8}) and (\ref{eq9}) we get:

\bigskip

 \begin{equation}\label{eq10}\rho^{\rm{app}}_{\mathbf{0}} (\beta,\nu,\nu)= \rho^{\rm{app}}(\beta,\mu,\nu) - \rho_{c}(\beta,\mu),\end{equation}

\bigskip

 \begin{equation}\label{eq11}\rho^{\rm{id}}_{\mathbf{0},l} (\beta,\mu,\nu) = \rho^{\rm{id}}_l (\beta,\mu,\nu) - \rho_{c,l}(\beta,\mu),\end{equation} being, as before: 

\bigskip

$$  \rho_{c,l} (\beta,  \mu )=  \frac{1}{V}  \displaystyle\sum_{\mathbf{p}\in {\Lambda_l}^*\backslash\{0\}} \left(\exp \beta \left( \frac{\mathbf{p}^2}{2} - \mu  \right)-1\right)^{-1},$$ 

\bigskip

$$
 \rho_c (\beta, \mu )=   \frac{1}{(2\pi)^d}  \int \left(\exp \beta \left(\frac{\mathbf{p}^2}{2} -  \mu \right)-1\right)^{-1}d^3\mathbf{p}.
$$

Since,

\[ \displaystyle\lim_{V\to\infty} \frac{1}{\beta V}\left( \frac{1}{e^{-\beta \mu} - 1}\right) = 0,\]

\bigskip

\noindent
for the fixed parameters $ (\beta,\mu ) \in \mathcal{D},\;\nu > 0,$  and from the expresions in eqs.(\ref{eq10}) and (\ref{eq11}), we have that both systems, simultaneously, undergo non conventional condensation. Moreover, the amount of condensate is given as:

\bigskip

\[ \rho^{\rm{app}}_{\mathbf{0}} (\mu,\nu) =  \rho^{\rm{id}}_{\mathbf{0}} (\mu,\nu) =\frac{\nu^2}{\mu^2}. \] 

\bigskip

\noindent

\end{proof}

\bigskip

The Bogoliubov's approach considers a chemical potential $\mu_{*}$ such that $\mu_{*}= -\frac{\nu}{ \sqrt{\rho_{\mathbf{0}}} },$ being $\rho_{\mathbf{0}}$ a real and  strictly positive constant

On the other hand, unlike the system given by the Hamiltonian  in eq.(\ref{eq1}), the system whose energy operator is represented by eq.(\ref{eq2}) preserves the U(1) symmetry.
 
If $\rho^{{\rm id}}_{0,l}\left(\beta,\mu_l,\nu\right)=\rho_{0}={\rm constant}\neq0,\;\rho_{0}>0 , \mu_l <0, $ we have that:

$$\rho_{0}\sim\frac{1}{\ V\left( -\mu_{l}+\mu^{2}_{l}/2\right) }+\frac{\nu^{2}}{\mu^{2}_{l}}.$$

\bigskip

Thus, for values of $ \mu_l $  in a small neighborhood of zero, 
$$\beta V\rho_{0}\sim -\frac{1}{\mu_l}+\frac{\beta V\nu^{2}}{\mu^{2}_{l}}.$$

By solving the second order equation in $ \mu_l,$ we obtain:
$$\mu_{l}\sim -\frac{1}{2\beta V\rho_{0}}\left( 1+\sqrt{1+\left( 2\beta V\nu\right) ^{2}\rho_{0}}\right) .$$

Finally, taking the  thermodynamic limit:
$$\displaystyle\lim_{V\to\infty}\mu_{l}=\mu^{\ast}=-\frac{\nu}{\sqrt{\rho}_{0}}. $$

For the free Bose gas, this result means that in the domain $\mathcal{D},$ in spite of that the chemical potential  $\mu_l$ depends on the inverse temperature $ \beta$ at finite volume, in the  thermodynamic limit   $\mu$ dependes only on the fixed parameters   $\rho_{0}, \nu.$

\bigskip

\subsection{Full diagonal models}

Let $\hat{H}^{\rm{FD}}_{l}(\mu) $ be the energy operator defined as:

\bigskip

\begin{equation}\label{eq12}\hat{H}^{\rm{FD}}_{l}(\mu) = \hat{H}^{0}_l+ \frac{a}{2V}\left(\hat{N}^2 -\hat{N}\right) + \frac{1}{2V}\displaystyle\sum_{\mathbf{p},\mathbf{p}^{\prime}}v(\mathbf{p}-\mathbf{p}^{\prime})\hat{n}_{\mathbf{p}}\hat{n}_{\mathbf{p}^{\prime}}.\end{equation}

\bigskip

$\hat{H}^{\rm{FD}}_{l}(\mu)$ belongs to a class  of energy operators so-called \emph{full diagonal Bose Hamiltonians}. Clearly, $\hat{H}^{\rm{FD}}_{l}(\mu)$ satisfies the commutation rule $[\hat{H}^{\rm{FD}}_{l}(\mu),\hat{N}]= 0.$ For example,  $\hat{H}^{0}_l$  is a full diagonal mode. If  $a >0, $  and  $v(\mathbf{p}-\mathbf{p}^{\prime})\geq 0$ these are superstable systems, i.e., their limit pressures exist for all real value of $\mu.$

\bigskip

Let $\hat{H}^{\rm{FD}}_{l,\nu}(\mu) ,\; \hat{H}^{\rm{FD, app}}_{l,\nu}(\mu)$ be the following operators:

\bigskip

\begin{equation}\label{eq13}\hat{H}^{\rm{FD}}_{l,\nu}(\mu) = \hat{H}^{\rm{FD}}_{l}(\mu) - \nu\sqrt{V} (\hat{a}^\dag_{\mathbf{0}} +   \hat{a}_{\mathbf{0}}),\;\nu > 0,\end{equation}

\bigskip

\begin{equation}\label{eq14}\hat{H}^{\rm{FD, app}}_{l,\nu}(\mu) = \hat{H}^{\rm{FD}}_{l}(\mu) - 2 \nu\sqrt{V} \sqrt{\hat{n}_{\mathbf{0}}+1},\;\nu > 0.\end{equation} 

\bigskip

In this case, $[\hat{H}^{\rm{FD, app}}_{l,\nu}(\mu), \hat{N}] =0, $ and  $[\hat{H}^{\rm{FD}}_{l,\nu}(\mu), \hat{N}] \neq 0. $ 

\bigskip

\begin{theorem}\label{thm3} 

\bigskip

\begin{equation}\label{eq15} p^{\rm{FD}} (\beta,\mu,\nu)  = p^{\rm{FD, app}}  (\beta,\mu,\nu). \end{equation}

\end{theorem}

\bigskip

\begin{proof}

\bigskip

For this kind of models in ref.~\cite{SUT} it has been proved that:

\begin{equation}\label{eq16}\displaystyle\lim_{V\to\infty}  \left\langle  \frac{\hat{a}^{\dag}_{\mathbf{0}}}{\sqrt{V}} \right\rangle_{\hat{H}^{\rm{FD}}
_{l,\nu}(\mu)} = \displaystyle\lim_{V\to\infty} \left\langle  \frac{\hat{a}_{\mathbf{0}}}{\sqrt{V}} \right\rangle_{\hat{H}^{\rm{FD}}
_{l,\nu}(\mu)}\end{equation} $$ = \rm{sgn\;\nu}\displaystyle\lim_{V\to\infty}\sqrt{V^{-1}\left\langle\hat{a}^{\dag}_{\mathbf{0}}\hat{a}_{\mathbf{0}} \right\rangle_{\hat{H}^{\rm{FD}}
_{l,\nu}(\mu)}      }. 
$$ 

\bigskip

\noindent
 In our case ${\rm sgn}\; \nu = + .$  

\bigskip

Let  define $\delta_{p_l},$ y $ \delta_H $ as 

$$\delta_{p_l}= p^{\rm{FD}}_l (\beta,\mu,\nu) - p^{\rm{FD, app}}_l (\beta,\mu,\nu),$$ 

$$ \delta_H = \hat{H}^{\rm{FD, app}}_{l,\nu}(\mu) - \hat{H}^{\rm{FD}}_{l,\nu}(\mu) = \nu\sqrt{V}\left( 2\sqrt{\hat{n}_{\mathbf{0}}+1}- (\hat{a}^{\dag}_{\mathbf{0}} + \hat{a}_{\mathbf{0}})\right),$$ 

\bigskip

\noindent
respectively. 

Note that $\hat{H}^{\rm{FD, app}}(\mu)$ preserves the $U(1)$ symmetry. This fact and the left hand side Bogoliubov's inequality (see the Apendix)  lead to:

\bigskip

$$ \delta p_l  \geq  2\nu\left\langle \sqrt{\hat{\rho}_{\mathbf{0},l}+\frac{1}{V}} \right\rangle_{\hat{H}^{\rm{FD, app}}
_{l,\nu}(\mu)}\geq 0. $$ Moreover, in the thermodynamic limit we get,

\bigskip

\begin{equation}\label{eq17}
\displaystyle\lim_{V\to\infty} \delta_{p_l} =  \geq  2\nu\displaystyle\lim_{V\to\infty}\left\langle \sqrt{\hat{\rho}_{\mathbf{0},l}} \right\rangle_{\hat{H}^{\rm{FD, app}}
_{l, \nu}(\mu)}\geq 0.
\end{equation}

\bigskip

From the right hand Bogoliubov's inequality and the Jensen inequa\-lity  (see Apendix) we obtain:

\bigskip

$$
\delta p_l \leq \frac{\nu}{\sqrt{V}}\left\langle  2\sqrt{\hat{n}_{\mathbf{0}}+1}- (\hat{a}^{\dag}_{\mathbf{0}} + \hat{a}_{\mathbf{0}}) \right\rangle_{\hat{H}^{\rm{FD}}
_{l,\nu}(\mu)} $$

 \begin{equation}\label{eq18} \leq  \nu \left(2\sqrt{\hat{\rho}_{\mathbf{0},l}  + \frac{1}{V}} - \left\langle  \frac{\hat{a}^{\dag}_{\mathbf{0}}}{\sqrt{V}} \right\rangle_{\hat{H}^{\rm{FD}}
_{l,\nu}(\mu)} - \left\langle\frac{\hat{a}_{\mathbf{0}}}{\sqrt{V}} \right\rangle_{\hat{H}^{\rm{FD}}
_{l,\nu}(\mu)}\right).
\end{equation}

\bigskip

Finally,  taking the limit $ V\to \infty$ and using the expresions in eq.(\ref{eq16}) and the inequalities (\ref{eq17}) and (\ref{eq18}),  we obtain: $ 0 \leq\displaystyle\lim_{V\to\infty}\delta p_l\leq 0. $
Hence $$p^{\rm{FD}} (\beta,\mu,\nu)  = p^{\rm{FD, app}}  (\beta,\mu,\nu).$$

\end{proof}

\bigskip

A well-known example of a full diagonal Hamiltonian is asociated to the mean field model, whose energy operato,r with an additional term broken the $ U(1)$  symmetry, is given by the expression: 

$$ \hat{H}^{\rm{MF}}_{l,\nu} = \hat{H}^{0}+ \frac{a}{2V}\left(\hat{N}^2 -\hat{N}\right) - \nu\sqrt{V} (\hat{a}^{\dag}_{\mathbf{0}}+\hat{a}_{\mathbf{0}}) ,$$ 

\bigskip

\noindent
where $a >0,$ $V$ is the volume of the region enclosing the particle system and $\nu\in\mathbb{R}.$ 

In this case, the operator $\hat{H}^{\rm{MF, app}}_{l,\nu}$ has the following form:

\bigskip

$$ \hat{H}^{\rm{MF, app}}_{l,\nu} = \hat{H}^{0}+ \frac{a}{2V}\left(\hat{N}^2 - \hat{N}\right) - \nu\sqrt{V} \sqrt{\hat{n}_{\mathbf{0}}+1}.$$ 

As before, $\nu >0.$

\bigskip

\subsection{Conclusions}

\bigskip

\begin{itemize}

\item[a.] For fixed parameters $\mu < 0,$ $\nu >0,$  the  pressures and the density of particles in the condensates of the systems whose operators are given by eqs.(\ref{eq1}) and (\ref{eq2}),  in the thermodynamic limit,  coincide. Thus,

\bigskip

\begin{equation*}  p^{\rm{app}}(\beta,\mu,\nu)  = p^{\rm{id}}(\beta,\mu,\nu)  = -\frac{\nu^2}{\mu} +   p^{\rm{id}^{\prime}} (\beta,\mu), \end{equation*}

\begin{equation*}    \rho^{\rm{id}}_{\mathbf{0}} (\mu,\nu) =
 \rho^{\rm{app}}_{\mathbf{0}} (\mu,\nu) = \frac{\nu^2}{\mu^2}, \end{equation*} 

\bigskip

\noindent
i.e., both models are equivalent in a thermodynamic sense and they undergo, simultaneously,  non conventional BEC  in $\mathcal{D}.$

\bigskip

\item[b.] The full diagonal models, with coupled external sources given in eqs. (13) and (14), are thermodynamically equivalent. 

\bigskip

\item[c.] Despite what has been said in a) and b), the  external source  $- 2 \nu\sqrt{V} \sqrt{\hat{n}_{\mathbf{0}}+1}$ does not remove the degeneracy of the 
regular averages.

\end{itemize}

\bigskip

\section{Apendix }

\bigskip

\subsection{Laplace principle}

\begin{proposition}\label{thm4}  Let $G:I\rightarrow R$ be a continuous function defined on the  interval  
 $I,$ and  bounded  above by the constant $M$ for all $x\in I$. It is assumed that  there exists $\alpha >0$ such that
for $|x|$ large enough,
\begin{equation*}
G(x)<-\alpha |x|.
\end{equation*}
Then,
\begin{equation}\label{eq19}
\lim_{N\rightarrow \infty }\frac{1}{N}\ln
\left(\int_{I}e^{NG(x)}dx\right)=\displaystyle\sup_{x\in I}\{G(x)\}.
\end{equation}
\end{proposition}

\subsection{Bogoliubov's Inequalities}

\bigskip

Let $\hat{H}_{a,l}$ and $\hat{H}_{b,l}$ be selfadjoint operators defined on $%
\mathcal{D} \subset {\mathcal{F}}_B$.  $p_{a,l}(\beta,\mu), $ $%
p_{b,l}(\beta,\mu) $ represent the grand canonical pressures and the free
canonical energies corresponding to the operators  $\hat{H}_{a,l}$, $\hat{H}%
_{b,l}$. In this case the following well known Bogolubov
inequalities,

\begin{equation}\label{eq20}  
\langle\frac{\hat{H}_{a,l}(\mu)-\hat{H}_{b,l}(\mu)}{V}\rangle_{\hat{H}%
_{a,l}(\mu)} \leq p_{b,l} (\beta,\mu) - p_{a,l} (\beta,\mu) \leq \langle%
\frac{\hat{H}_{a,l}(\mu) - \hat{H}_{b,l}(\mu)}{V}\rangle_{\hat{H}%
_{b,l}(\mu) },
\end{equation}
hold, where $\langle - \rangle_{\hat{H}_{a,l}(\mu)},$ $\langle - \rangle_{%
\hat{H}_{b,l}(\mu)}$  are the Gibbs states in the grand canonical ensemble
associated to the Hamiltonians $\hat{H}_{a,l}, \hat{H}_{b,l}, $ respectively.

\bigskip

\subsection{Jensen's Inequality}

Let $\hat{H}_l,$ be a self-adjoint operator,  diagonal with respect to the number operators.Since the spectrum of  $\hat{H}_l,$ coincides with the
set of non negative integers,
 this model can be classically understood
 by using non negative random variables defined on a suitable probability space $\Omega_l.$  
 
Let $\Omega_l$ be the countable  set of sequences $\omega = \{\omega (p)\in {\mathbb N}: p\in
\Lambda^*_l\}\subset
 {\mathbb N} \cup \{0\}$ satisfying
\begin{equation*}
\sum_{p\in \Lambda^*_l}\omega (p)<\infty \;.
\end{equation*}
The basic random variables are the occupation numbers
$\{n_{p}:j=1,2,....\}$. They are defined as the
functions $ n_{p}:\Omega _{l}\rightarrow {\mathbb{N}}$ given
as $n_{p}(\omega )=\omega (p)$ for any $\omega \in
\Omega _{l}$. The total number of particles in the configuration
$\omega $ is denoted as $N(\omega ).$ Then the total number,
excluded the zero mode is denoted as $N^{\prime}(\omega ).$

\bigskip

In this framework, the Gibbs state can be written by replacing $\hat{H}_l,$ by a function  $H_l: \Omega_l \rightarrow {\mathbb R},$ representing the proyection of the energy operator on the occupation-number basis of the Bose Fock space.

\bigskip

Let ${\mathbb P}$ be a probability defined for any
$\omega\in\Omega_l$ as

\begin{equation}\label{eq21}{\mathbb P}[\omega] =  \left[
\displaystyle\sum_{\omega\in\Omega_l}\exp \left( -\beta [H_l
(\mu)](\omega)\right)\right]^{-1}\exp \left( -\beta [H_l
(\mu)](\omega)\right).\end{equation} For arbitrary $S\subset\Omega$
this implies that
\begin{equation}
{\mathbb P}[S\subset\Omega_l] =  \left[
\displaystyle\sum_{\omega\in\Omega_l}\exp \left( -\beta [H_l
(\mu)](\omega)\right)\right]^{-1}\displaystyle\sum_{\omega\in S}\exp
\left( -\beta [H_l (\mu)](\omega)\right).
\end{equation}

\bigskip

In this case, $\left<  \hat{X} \right>_{\hat{H}_l (\mu)} \equiv {\mathbb E}[X],$ being $X: \Omega_l \rightarrow \mathbb{R}$ the function corresponding to the proyection of the operator $\hat{X}$ on the occupation-number basis.

Thus, the expectation of $X$ respect to ${\mathbb P} $  is defined as:

\begin{equation}\label{eq22}
{\mathbb E}[X] =  \displaystyle\sum_{\omega\in \Omega_l} X(\omega) {\mathbb P} [\omega].
\end{equation}

\bigskip

If $X: \mathbb{R}\rightarrow \mathbb{R}$ is a concave function, the following Jensen's inequality:

\begin{equation} {\mathbb E}[f(X)] \leq f({\mathbb E}[X]),\end{equation} holds.

\bigskip

\subsection{Griffiths Lemma}

\bigskip

\begin{Lemma}\label{thm5} \mbox{(Griffiths \cite{GRI})} Let $\{ g_n: I  \to \mathbb{R},\; I \equiv (a,b) \subset \mathbb{R}\}_{n\in \mathbb{N}} $ be a sequence of convex functions on $I$ with a pointwise limit $g(x),$ which, of course is convex. Let $G^{+}_n (x)$ $[ \mbox{resp.} \;G^{-}_n (x) ]$ be the right (resp. left) derivatives of $g_n(x),$ and similarly for $G^{+}(x),\;G^{-} (x).$ Then, for all $x\in I,$

\begin{equation}
\displaystyle\lim_{n\to\infty} \sup G^{+}_n (x) \leq  G^{+} (x), \;\; \displaystyle\lim_{n\to\infty} \sup G^{-}_n (x) \geq  G^{-} (x).
\end{equation}

In particular, if all the $g_n$ and $g$ are differentiable at some point $x\in I,$ then

\begin{equation}
\displaystyle\lim_{n\to\infty}\frac{dg_n (x)}{dx} = \frac{dg (x)}{dx}.
\end{equation} 
\end{Lemma}

\bigskip

\begin{proof} Fix $x\in I$ and $ x\pm y \in I,$

\[\ g_n (x+y) \geq g_n(x) + yG^{+}_n (x), \]

\[g_n (x-y) \geq g_n(x) - yG^{-}_n (x). \]

\bigskip

Fix $y$ and take the limit $n\to \infty.$ Then,

\[\displaystyle\lim_{n\to\infty} \sup G^{+}_n (x) \leq y^{-1}[g(x+y) - g(y)] \] 

\bigskip

\noindent
and similarly for $\displaystyle\lim_{n\to\infty} \inf  G^{-}_n (x).$ Now let $ y\downarrow 0.$

\end{proof}

\end{document}